\documentclass[]{llncs}
\usepackage[T1]{fontenc}
\usepackage{graphicx}
\usepackage{cite}
\usepackage{booktabs}

\begin{document}
\title{A Computer-Supported Collaborative Learning Environment for Computer Science Education}
\titlerunning{A Computer-Supported Collaborative Learning Environment for CSE}
%
\author{Michael Holly\inst{1} \and
Jannik Hildebrandt\inst{1} \and
Johanna Pirker\inst{1,2}}
\authorrunning{M. Holly et al.}
%
\institute{Graz University of Technology, Austria \and
Ludwig-Maximilians-Universität München, Germany
\email{\{michael.holly,johanna.pirker\}@tugraz.at}\\
\email{jhildebrandt@student.tugraz.at}
}
\maketitle              
\begin{abstract}
Skills in the field of computer science (CS) are increasingly in demand. Often traditional teaching approaches are not sufficient to teach complex computational concepts. Interactive and digital learning experiences have been shown as valuable tools to support learners in understanding. However, the missing social interaction affects the quality of the learning experience. Adding collaborative and competitive elements can make the virtual learning environment even more social, engaging, and motivating for learners.
In this paper, we explore the potential of collaborative and competitive elements in an interactive virtual laboratory environment with a focus on computer science education. In an AB study with 35 CS students, we investigated the effectiveness of collaborative and competitive elements in a virtual laboratory using interactive visualizations of sorting algorithms.


\end{abstract}
\section{Introduction}

Today, innovative technologies play an evolving role in our everyday lives. Science, technology, engineering, and mathematics (STEM) are becoming more relevant. The digital transformation in the industry made computer science (CS) an important field. Therefore, computer science education (CSE) is an essential element in addressing the lack of experts in this field. Yadav et al. \cite{yadav2016} demonstrated already the challenging nature of teaching CS topics. Traditional teaching methods present solutions and concepts, but they fail in teaching problem-solving.
In contrast, integrated educational activities that engage learners in the learning process have proven to be a successful teaching method. These learning tools help students to improve their understanding of conceptual aspects. Two of these methods are digital learning and collaborative learning. Digital learning integrates computers or other technologies into the learning process, which allows students to learn from home. However, this approach misses the social component by eliminating direct communication with others \cite{wheeler2005creating}. At this point, collaborative learning comes into play. It takes advantage of going through the learning process as part of a group where students benefit from each other's knowledge and experience. Although these two methods are different, they can be combined to form computer-supported collaborative learning environments. This approach allows students to work together in a digital world and combines the advantages of both methods. Nevertheless, there are only a few examples that implemented this method in large-scale interactive learning environments, and there are even fewer that have accomplished collaboration via the Internet.

In this paper, we want to introduce a computer-supported collaborative environment for CSE integrated into a virtual laboratory where the users can work together on different experiments.
The main research objectives are:
\begin{itemize}
    \item Exploring how collaboration affects learners’ motivation, emotions, and learning outcomes.
    \item Investigating the connection between team partners and the effect on learning and engagement in a collaborative environment.
    \item Identifying the user acceptance of a battle mode with competitive elements in a collaborative environment.
\end{itemize}

\paragraph*{Contribution}
In this paper, we present a study with 35 CS students, discussing a computer-based collaborative laboratory environment for computer science education. The focus is on identifying and discussing the benefits and challenges of conceptual learning in a collaborative virtual environment with the target group: CS students.
\section{Background and Related Work}

Traditional teaching approaches for CSE often involve only didactic instructions. Studies show that teaching methods that are more engaging and interactive can enhance these learning methods \cite{aycock2019adapting, pirker2014motivational}. Peters \cite{Peters_2000} analyzed such digital learning environments from a pedagogical perspective and pointed out that they are becoming more open, flexible, and variable in teaching and learning. It allows to adapt it to learners' needs and increases the motivation and also the time that students spend with the learning material. Moreover, it allows students to learn from anywhere and at any time. Papastergiou \cite{papastergiou2009digital} investigated the learning effectiveness and motivational appeal of a digital game-based learning approach for computer memory concepts. The study demonstrated that this approach was both more effective in promoting students’ knowledge of computer memory concepts and more motivational. While out-of-school learning is becoming more powerful and popular, studies have shown that in-school learning is still critical for the learning outcome. Warschauer \cite{warschauer2007paradoxical} points to the role of social, cultural, and economic factors in shaping and constraining educational transformation in the digital area.
Through communication and social interaction, virtual worlds are an ideal platform for engaging learners in educational settings \cite{Moschini2010}. Gütl \cite{Guetl2011} describes it as a potential way to mitigate or even overcome collaboration issues in existing technologies. Virtual worlds provide a set of tools to foster effective group collaboration for different digital learning scenarios. The use of avatars, the support of verbal and non-verbal communication, and creative capabilities are key elements for effective group learning in virtual worlds \cite{franceschi2008engaging}.
Crellin et al. \cite{crellin2009virtual} showed that this can also be used in different CSE areas as a development environment, a collaboration tool, or to provide an environment for simulations. Cerny and Mannova \cite{cerny2011competitive} demonstrated a competitive and collaborative approach to make learning in computer science more effective.
In CS, many efforts have been made to support students in learning through computer-supported techniques such as visualizations and simulations. This includes topics such as computer networking, software engineering, computer architecture, or computer science principles \cite{alnoukari2013simulation, wolff2000satsim}.
The Computer Science Unplugged project explores different approaches to teach children math and computing topics through unplugged activities. They demonstrated this concept in a parallel sorting network where the students had to work together to get to the other side of the network in the correct order. This approach was also transferred into a virtual environment to teach the concepts to those who are unable to participate physically \cite{bell2009computer}. SATSim goes one step further and provides an animated and interactive visualization aid for teaching superscalar architecture concepts. It includes out-of-order execution, in-order commitment, dynamic resolution of data dependencies using register renaming and reservation stations, and the performance effects of branch prediction accuracy and cache hit rates. The concept was included in an advanced undergraduate computer architecture course to visualize the complicated behavioral patterns of superscalar architectures. The study results indicated that there is a significant improvement in students' understanding when using animated and interactive visualizations \cite{wolff2000satsim}. Moreover, practical experiences are essential in understanding and handling software issues. The SimSE environment is an interactive simulation game for software engineering education that allows students to take on the role of a project manager to deal with a specific situation that arises during a software engineering process \cite{navarro2004simse}. They showed in a multi-angled evaluation approach that students can learn the concepts successfully presented in an enjoyable experience but mentioned that it is most effective when used as a complementary component to other teaching methods \cite{navarro2007comprehensive}.

While many studies show a positive effect on learning, engagement, and motivation, it is crucial to understand better the potential of collaborative and competitive elements in virtual environments for CSE.
\section{Learning Application}
The virtual learning application provides an immersive 3D laboratory and experiment environment that allows users to learn different phenomena by conducting interactive experiments. The laboratory is designed as a modular extendable framework where experiments can be independently added to the learning environment. A lobby room acts as a three-dimensional menu that displays the different stations of the available experiments. The stations themselves are entry points that allow the user to access the learning activities. The desktop version allows the user to control the application using the keyboard and mouse, similar to a classic computer game. When the user enters an experiment, a stand-alone scene is loaded in which the user can experience the interactive simulation. All learning activities and experiments are designed for active learning and support several virtual learning experiences with different forms of engagement and immersion. Platform-specific virtual control elements allow users to modify several experiment parameters to demonstrate the effect of these parameters on the experiment outcome \cite{pirker2017improving}.

\subsection{Collaborative Experiment Setup}

Based on the desktop version, a multi-user network was added. It allows users to work together on different experiments. This form of learning has great potential, especially in virtual learning environments. Through the integration of social interactions, the engagement of the users should be deepened. For this purpose, we developed a network manager that extends the laboratory environment with server-client communication and synchronization. When joining the lab lobby room, users can create a server or connect to an existing one, either over a local network or over the Internet. When multiple users are connected, there is always one user in control who can enter experiments and control the experiment parameters. These user actions are then distributed over the network so that everyone in the network has the same experiment state and can discuss it together with the others. The other users can request the control at any time to affect the experiment as well. The communication during the experiment is done via voice calls, video calls, or text messages using an external tool such as Discord\footnote{https://discord.com/}. It allows both use cases where a teacher demonstrates an experiment to the class and where they work together collaboratively. After a guided session, the control can be released to allow students to explore the experiment themselves at their own pace.

\subsection{Experiments and Simulations}

The laboratory contains nine experiments: Seven on physics, one on chemistry, and one on computer science. In this paper, we focus only on the computer science experiment "Sorting Algorithms" to identify the benefits and challenges of a collaborative environment for conceptual learning. The other experiments were excluded from the study. The goal of the CS simulation is to demonstrate and visualize the concept of multiple sorting algorithms using two different views: (1) a detail-view and (2) a battle-view.

\begin{figure*}
\centerline{\includegraphics[width=26pc]{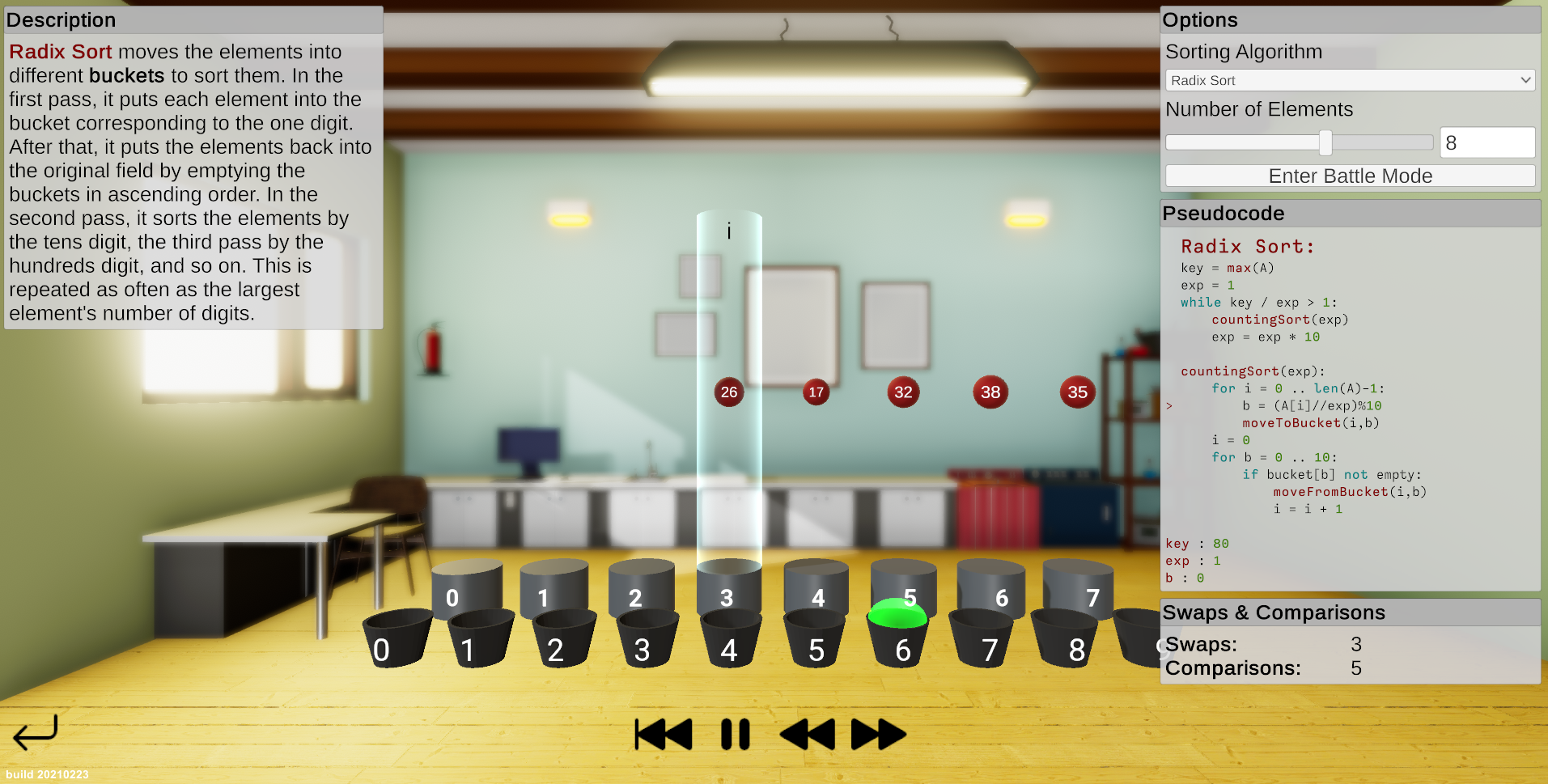}}
\caption{The \textsc{detail-view} of the experiment shows the sorting visualization including the current execution line and a description of the loaded algorithm.}
\label{fig:DetailView}
\end{figure*}

The \textsc{detail-view} (see {Fig. \ref{fig:DetailView}}) allows users to investigate nine different algorithms by stepping through the algorithm forward and backward. The sorting field is visualized by different-sized spheres with numbers to be sorted in ascending order. During the operations, a highlighted pseudo-code illustrates the algorithm and shows the current line of execution. Additionally, a short textual description of the selected algorithm is displayed to explain the idea of the sorting algorithm. 

\begin{figure*}
\centerline{\includegraphics[width=26pc]{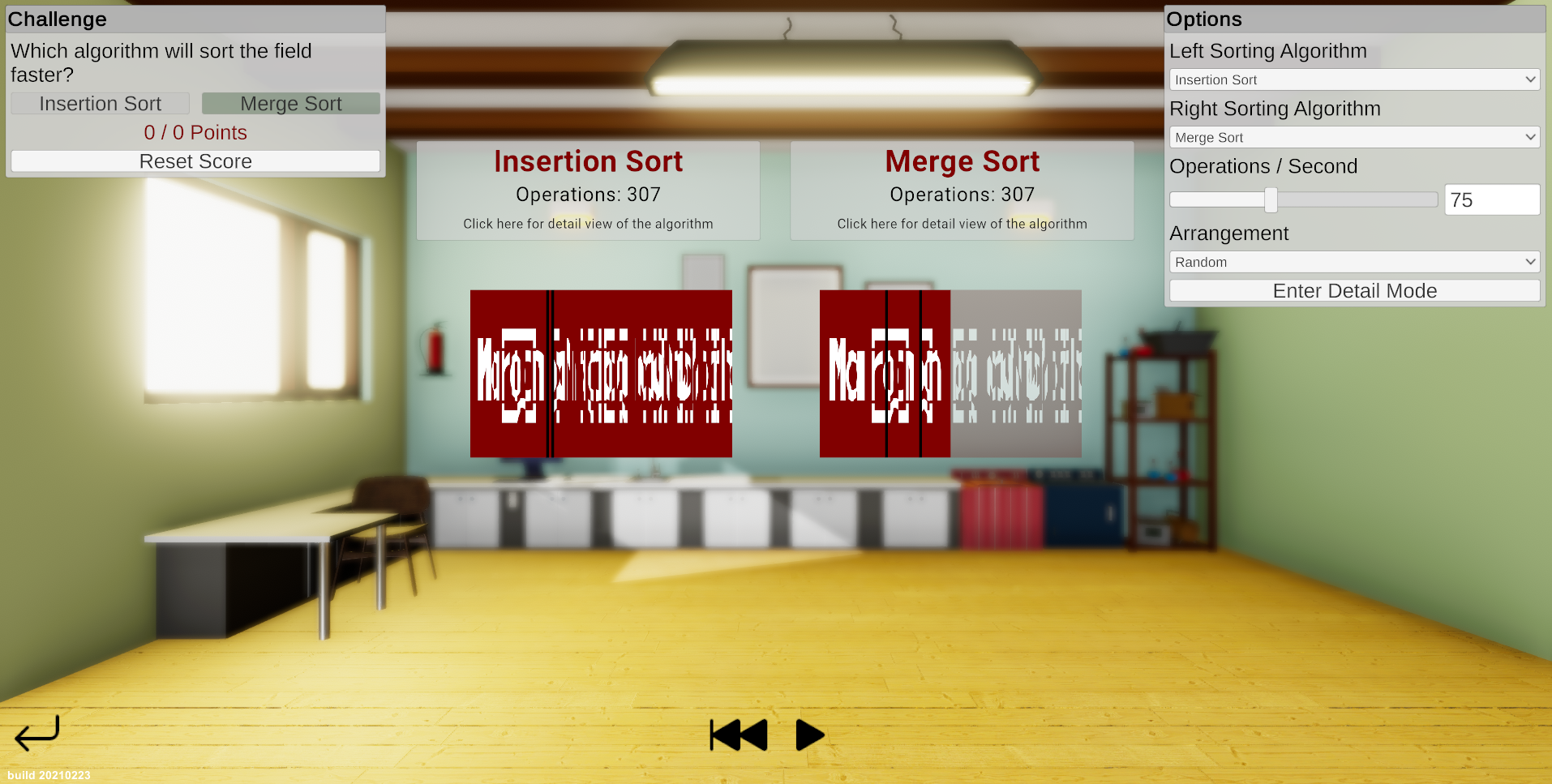}}
\caption{The \textsc{battle-view} of the sorting experiment allows the comparison of two different algorithms on a given input array to evaluate their efficiency.}
\label{fig:BattleView}
\end{figure*}

In the \textsc{battle-view} (see {Fig. \ref{fig:BattleView}}), users can pit algorithms against each other for a better understanding of the efficiency of the algorithms.
For this purpose, an image is divided into 100 stripes representing the elements to be sorted into the correct order. Users can select if the elements should be arranged randomly, reversed, or already sorted. The challenging part for the user is to guess in advance which algorithm will sort the field faster. For each correct guess, they receive a point. As a particular highlight, each user can participate in the challenge individually, with a scoreboard keeping track of the points. This adds a competitive spirit to the collaborative environment.

\section{Evaluation}

In previous studies, we focused on different learning experiences in terms of immersion, engagement, usability, and user experience. However, these studies missed the collaborative learning aspects. Therefore, in this paper, we focus on identifying the benefits and challenges of conceptual learning in a collaborative environment. We conducted an AB study with CS students (target group) to compare the multi-user version of the sorting experiment with the single-user version. For this purpose, we built upon the framework by Naps et al. \cite{naps2002} which provides a basis to measure the effectiveness of algorithm visualizations. The design of the study was constructed with a focus on experience and engagement, learning outcomes, and usability.

\subsection{Material and Setup}

The study was conducted during the COVID-19 pandemic. Participants were separated from each other and located in different places at home. All participants used their personal computers connected to the Internet.  The 35 participants were randomly assigned to the multi-user group (20 participants) or the single-user group (15 participants). In the multi-user group, participants worked in pairs, allowing collaboration, whereas the single-users worked autonomously. The communication with the participants and between the multi-user group members was done via Discord.

\subsection{Method and Procedure}

The study consists of a pre-questionnaire, the actual tasks in the learning application, and a post-questionnaire.
In the beginning, we asked all participants to fill out a pre-questionnaire to gather previous experiences with digital learning, sorting algorithms, and their programming skills. They were asked to answer ten theoretical questions on sorting algorithms based on Bloom's taxonomy\footnote{https://bloomstaxonomy.net/}, targeting different levels of understanding. Each of these questions was graded by computer scientists (author of the paper) as either 0, 0.5, or 1 point, depending on how accurately the question was answered, leading to a maximum of 10 points.
After completing the pre-questionnaires, they got a detailed introduction to the test system. We explained to them how to move and interact with the virtual environment. The multi-users were then asked to join a voice chat with their partner. In the laboratory, they joined an existing server and were asked to perform a series of tasks together. The single users completed the same tasks autonomously. Both groups had to complete the following tasks:

\begin{itemize}
    \item Familiarize yourselves with merge sort, insertion sort, and radix sort without using the battle-view mode. There will be a small challenge at the end, where you can apply your knowledge.
    \item Once you feel confident that you understand the sorting algorithms, switch to the battle-view. Before you start sorting, always choose one of the two algorithms in the challenge that you think will perform faster.
    \begin{itemize}
        \item Let merge sort compete against insertion sort.
        \item Let merge sort compete against radix sort.
        \item Change the arrangement to sorted. Then let insertion sort compete against radix sort.
    \end{itemize}
\end{itemize}
When finished with the tasks, the multi-users had to leave the voice channel. After conducting the experiment, all participants were asked to fill out a post-questionnaire. They had to answer the same ten theory questions in a randomized order to check their conceptual understanding. The participants were also asked to answer open-ended questions and 6 questions on a Likert scale between 1 (not at all) and 5 (very much) about their overall experience and the integrated battle mode acceptance. To measure the motivation and learning experience towards the simulation, we asked them to fill out 16 questions on a Likert scale between 1 (fully disagree) and 7 (fully agree). We used the System Usability Scale (SUS) \cite{brooke2013sus} to measure the system usability and the Computer Emotion Scale (CES) \cite{kay2008assessing} to evaluate the users' emotions while interacting and learning with the virtual environment. To investigate the connection between the multi-user pairs, we selected relevant questions of the Classroom Community Scale \cite{rovai2002development} and asked them to answer the Online Student Engagement questionnaire \cite{dixson2015measuring} to get insights into their collaborative learning habits.

\subsection{Participants}

The study was conducted with 35 participants (29 male; 6 female) with a background in computer science. To recruit the participants, we contacted students via social media channels. They were aged between 21 and 34 (AVG=26.91; SD=2.98). In the pre-questionnaire, we asked each of them to rate their experience with computers, video games, programming, and sorting algorithms on a Likert scale from 1 (low) to 5 (high). Most of the participants rated themselves as an expert in computer usage (SU: AVG=3.93, SD=1.22; MU: AVG=4.1; SD=1.12), video games (SU: AVG=3.67, SD=1.23; MU: AVG=2.85, SD=1.27), and programming (SU: AVG=3.87, SD=1.46; MU: AVG=3.20, SD=1.24). Participants indicated that they are familiar with sorting algorithms (SU: AVG=2.73, SD=0.80; MU: AVG=2.35, SD=0.88). 15 participants had already used an e-learning tool.
 \section{Results}
The following section presents the results of the single-user group (SU) and the multi-user group (MU) with a focus on collaboration, online engagement, and learning outcomes. Since the learning outcome depends on the user experience and the system acceptance, we investigate also the system usability and the learning experience.

\subsection{Usability and User Experience}
Participants rated their overall impression and acceptance on a Likert scale from 1 (not at all) to 5 (very much) - {Table \ref{tab:OverallImpression}}. In general, all users found the sorting experiment interesting and enjoyable. For the battle view, there was a significant difference between the single-users (AVG=4.73, SD=0.46) and the multi-users (AVG=3.95, SD=1.10); Wilcoxon rank-sum test: $W = 217.5, p = 0.014$. However, they agreed that the battle view provides a clear understanding of the complexity of the algorithms. The additional multi-user features for the experiment were accepted by most users (MU: AVG=3.60, SD=1.23). Only four users disliked the concept that only one user had control over the experiment and would prefer more interactions for all clients. To measure the emotions of happiness, sadness, anger, and anxiety during the learning process, we used the Computer Emotion Scale. {Table \ref{tab:CES}} summarizes the results of the CES items for the single-user and the multi-user. Both groups rated happiness (e.g. satisfied, excited, curious) as high and the emotions of sadness, anger, and anxiety as very low. The only significant difference was that the single-users (AVG=0.87, SD=0.34) felt more insecure than the multi-users (AVG=0.53, SD=0.50); Wilcoxon rank-sum test: $W = 191, p = 0.04$. The overall application usability was evaluated with the SUS questionnaire. The single users rated the usability with a score of 81.67, which indicates good usability. In comparison, the multi-users scored usability slightly lower with 72.24. The two groups differed most significantly on whether the system was easy to use (SU: AVG=4.27, SD=0.70, MU: AVG=3.53, SD=9.96); Wilcoxon rank-sum test: $W = 204.5, p = 0.024$.

\begin{table*}
\centering
\caption{Overall Impression and Battle Mode Acceptance on a Likert Scale between 1 (not at all) and 5 (very much)}
\label{tab:OverallImpression}
\begin{tabular}{p{8.8cm}llll}
\toprule
& \multicolumn{2}{l}{Single-User} & \multicolumn{2}{l}{Multi-User} \\
& AVG            & SD             & AVG            & SD            \\
\midrule
I liked the sorting experiment.  & 4.33           & 0.62           & 4.15           & 0.81          \\
The challenge made the Battle Mode more interesting.   & 4.40           & 0.83           & 3.80           & 1.06          \\
Knowing that there is a challenge at the end motivated me to be more active.     & 3.87           & 0.92           & 3.70   &1.30\\
I liked the Battle Mode.   & 4.73           & 0.46           & 3.95           & 1.10          \\
The battle mode gave me a clearer understanding of the complexity of the algorithms.   & 4.33           & 0.72           & 4.00           & 1.30         \\
I would like to do the challenge in a bigger group, to compete with other people.   & 3.00           & 1.00           & 3.30           & 1.38          \\
\bottomrule
\end{tabular}
\end{table*}

\begin{table}
\centering
\caption{Computer Emotion Scale on a Likert Scale between 0 (never) and 3 (always)}
\label{tab:CES}
\begin{tabular}{lllll}
\toprule
          & \multicolumn{2}{l}{Single-User} & \multicolumn{2}{l}{Multi-User} \\
          & AVG            & SD             & AVG            & SD            \\
\midrule
Happiness & 1.69           & 0.68           & 1.89           & 0.74          \\
Sadness   & 0.23           & 0.39           & 0.21           & 0.47          \\
Anger     & 0.11           & 0.31           & 0.14           & 0.37          \\
Anxiety   & 0.33           & 0.35           & 0.24           & 0.38          \\
\bottomrule
\end{tabular}
\end{table}

\subsection{Learning Experience and Outcome}
\begin{figure}
\centerline{\includegraphics[width=18.5pc]{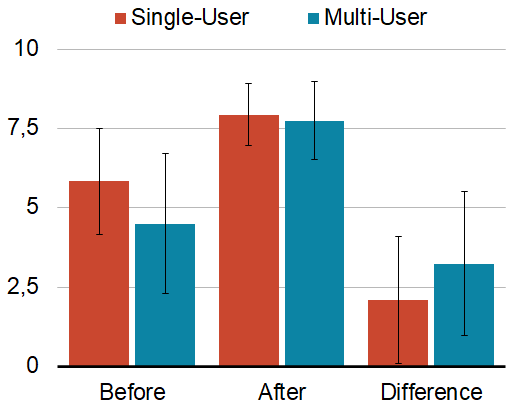}}
\caption{Results of the ten theory questions before and after the learning experience. Each question was graded with 0, 0.5, or 1 point, depending on how accurate the answer was.}
\label{fig:ResultLerningOutcome}
\end{figure}

To evaluate the learning experience, we asked the participants to rate their experience on a Likert scale between 1 (not agree) and 7 (fully agree). {Table \ref{tab:LearningExperience}} gives an overview of the users' learning experience. The responses regarding the learning experience were generally positive. Both groups agreed that they learned something from the experiment. They mentioned that the application is a good supplement for learning. Single users indicated that the experience was more engaging and fun. In contrast, multi-users were less engaged and had a reduced sense of fun. However, both groups reported that the experience inspired them to learn more about sorting algorithms (SU: AVG=4.73, SD=1.58; MU: AVG=5.21, SD=1.81). They would also prefer to learn at home (SU: AVG=5.87, SD=1.06; MU: AVG=4.42, SD=2.27) than in classrooms (SU: AVG=5.20, SD=1.93; MU: AVG=4.26, SD=2.02).
Before and after the learning session, the participants were asked several theoretical questions (max 10 points) to determine the learning outcome. In the pre-questionnaire, single users performed significantly better than the multi-users (SU: AVG=5.83, SD=1.67; MU: AVG=4.50, SD=2.22). After the learning experiment, both groups improved their knowledge, with the single-user group still slightly ahead (SU: AVG=7.93, SD=0.98; MU: AVG=7.74, SD=1.23). However, the multi-users showed a higher improvement in their knowledge than the single-users. The increased learning outcome was for both groups significantly higher; Wilcoxon rank-sum test: $p<0.001$. {Fig. \ref{fig:ResultLerningOutcome}} shows the user's learning performance before and after the experiment. The time spent in the detail-view before users switched to the battle-view varied for both groups. The single users spent 9.64 minutes on average, while the multi-users took 14.02 minutes. Nevertheless, both groups performed equally well on the tree tasks in the battle-view (SU: AVG=2.40, SD=0.63; MU: AVG=2.56, SD=0.51).

\begin{table*}
  \centering
  \caption{Learning Experience on a Likert Scale between 1 (fully disagree) and 7 (fully agree)}
  \label{tab:LearningExperience}
  \begin{tabular}{p{8.8cm} c c c c}
    \toprule
          & \multicolumn{2}{l}{Single-User} & \multicolumn{2}{l}{Multi-User} \\
          & AVG            & SD             & AVG            & SD            \\
    \midrule
I would   like to learn with it                                                                                         & 5,40 & 1,40 & 4,89 & 1,94 \\
It is a good idea to use it   for learning                                                                              & 6,40 & 0,83 & 5,21 & 1,78 \\
It is a good supplement to   regular learning                                                                           & 6,47 & 0,83 & 5,74 & 1,41 \\
I learned something with it                                                                                             & 6,07 & 1,39 & 6,11 & 1,52 \\
It makes the content more   interesting                                                                                 & 6,13 & 0,74 & 5,68 & 1,53 \\
It makes the content easier to   understand                                                                             & 6,33 & 0,98 & 5,42 & 1,39 \\
It makes learning more   engaging                                                                                       & 6,27 & 0,70 & 5,53 & 1,35 \\
It makes learning more fun                                                                                              & 6,13 & 0,99 & 5,63 & 1,50 \\
It makes learning more   interesting                                                                                    & 6,00 & 1,13 & 5,68 & 1,25 \\
I would like to learn with it   at home                                                                                 & 5,87 & 1,06 & 4,42 & 2,27 \\
I would like to learn with it   in the classroom                                                                        & 5,20 & 1,93 & 4,26 & 2,02 \\
The experience inspired me to   learn more about sorting algorithms                                                     & 4,73 & 1,58 & 5,21 & 1,81 \\
Learning was more motivating   than ordinary exercises                                                                  & 5,73 & 1,16 & 5,68 & 1,25 \\
I would rather like to learn   sorting algorithms with the sorting algorithm experiment than with   traditional methods & 6,20 & 1,32 & 5,42 & 1,68 \\
I find regular computer   science classes boring                                                                        & 2,93 & 1,53 & 2,16 & 1,12 \\
Seeing the sorting algorithm visualizations on the computer was engaging                                              & 6,20 & 1,08 & 5,79 & 1,13 \\

    \bottomrule
  \end{tabular}
\end{table*}

\subsection{Collaboration and Online Engagement}

To investigate the collaboration and the connection between the team partners in the multi-user group, we used the selected questions from the Classroom Community Scale questionnaire. The users rated their experience on a Likert scale from 0 (fully disagree) and 4 (fully agree). Users agreed that they could rely on their team partner (AVG=3.42, SD=0.67). Ten users even stated that they fully agreed with this statement, and none disagreed. Users did not feel uneasy about exposing their knowledge gaps (AVG=0.68, SD=0.92) or reluctant to speak openly (AVG=0.47, SD=0.82). They also disagreed with the statement that their team partner did not help them learn (AVG=0.42, SD=0.82). Since engagement can affect learning, especially in online scenarios where users often feel isolated and involved, we evaluated user engagement for both user groups. The participants were asked to rate the statements about engagement (Online Student Engagement questionnaire) on a Likert scale from 1 (not at all characteristic of me) to 5 (very characteristic of me). Users described themselves concerning certain behaviors, thoughts, and feelings. Multi-users had a slightly higher motivation to get a good grade (SU: AVG=3,93, SD=0.96; MU: AVG=3.95, SD=1.18) or to perform well on the test/quiz at the end (SU: AVG=3.93, SD=1.03; MU: AVG=4.00, SD=1.15). There was also a significant difference in whether they actively participated in small group discussion forums (SU: AVG=2.53, SD=3.63; MU: AVG=3.58, SD=1.17); Wilcoxon rank-sum test: $W = 79.5, p = 0.026$. In total, the multi-users ranked their engagement higher in 17 of 19 characteristics. Table \ref{tab:Collaboration} summarizes the collaboration scores, while Table \ref{tab:OSE} displays the users' online engagement scale for both groups.

\begin{table}[h]
  \centering
  \caption{Collaboration Scale on a Likert Scale between 0 (fully disagree) and 4 (fully agree)}
  \label{tab:Collaboration}
  \begin{tabular}{p{8.8cm} c c}
    \toprule
      & AVG & SD\\
    \midrule
    I feel that I am encouraged to ask questions & 3.16 & 1.14 \\
    I feel that it is hard to get help when I have a question & 0.95 & 1.19 \\
    I do not feel a spirit of community & 0.63 & 1.04 \\
    I feel uneasy exposing gaps in my understanding & 0.68 & 0.92 \\
    I feel reluctant to speak openly & 0.47 & 0.82 \\
    I feel that I can rely on my team partner & 3.42 & 0.67 \\
    I feel that my team partner does not help me learn & 0.42 & 0.82 \\
    I feel that I am given ample opportunities to learn & 3.11 & 0.79 \\
    I feel uncertain about my team partner & 0.37 & 0.93 \\
    \bottomrule
  \end{tabular}
\end{table}

\begin{table}[h]
  \centering
  \caption{Online Student Engagement Scale on a Likert Scale between 1 (not at all characteristic of me) and 5 (very characteristic of me)}
  \label{tab:OSE}
  \begin{tabular}{p{8.8cm} c c c c}
    \toprule
          & \multicolumn{2}{l}{Single-User} & \multicolumn{2}{l}{Multi-User} \\
          & AVG            & SD             & AVG            & SD            \\
    \midrule
    Making sure to study on a regular basis & 2.53 & 0.92 & 3.21 & 1.13 \\
    Putting forth effort & 3.40	& 0.99 & 3.47 & 1.31 \\
    Staying up on the readings & 2.80 &	1.01 & 2.79 & 0.92 \\
    Looking over class notes between getting online to make sure I understand the material & 2.87 &	1.13 & 2.58 & 0.96 \\
    Being organized & 3.27 &	1.22 & 3.47 & 1.26 \\
    Taking good notes over readings, PowerPoints, or video lectures & 2.73 &	1.10 & 2.79 & 1.32 \\
    Listening/reading carefully & 3.53	& 1.06 & 3.63 & 1.07 \\
    Finding ways to make the course material relevant to my life & 2.80	& 1.08 & 3.16 & 0.96 \\
    Applying course material to my life & 3.00 & 0.93 & 3.26 & 1.1 \\
    Finding ways to make the course interesting to me & 3.27 &	1.03 & 3.42 & 0.90 \\
    Really desiring to learn the material & 3.40 & 0.74 & 3.63 & 1.01 \\
    Having fun in online chats, discussions, or via email with the instructor or other students & 2.40 & 0.99 & 3.11 & 1.37 \\
    Participating actively in small-group discussion forums & 2.53 & 1.25 & 3.58 & 1.17 \\
    Helping fellow students & 4.07 & 0.80 & 4.16 & 0.96 \\
    Getting a good grade & 3.93 & 0.96 & 3.95 & 1.18 \\
    Doing well on the tests/quizzes & 3.93 & 1.03 & 4.00 & 1.15 \\
    Engaging in conversations online (chat, discussions, email) & 2.27 & 1.16 & 3.16 & 1.26 \\
    Posting in the discussion forum regularly & 1.80 & 0.86 & 2.63 & 1.30 \\
    Getting to know other students in the class & 3.27 & 1.33 & 3.84 & 1.21 \\
    \bottomrule
  \end{tabular}
\end{table}
\section{Discussion}

In this study, we focused on students' learning outcomes and experiences in a computer-supported collaborative environment for computer science education. We tried to investigate the effect of competitive elements in an online collaborative environment with the target group: CS students. 
The goal was to explore how collaboration affects learners' emotions and learning outcomes. The results showed that the multi-users improved their learning outcome significantly higher than the single-user group. This fact is relativized by the performance of the single users, who performed better in total. Also, the different group achievement levels at the beginning have to be considered. For students with a higher level, it is more difficult to improve their knowledge. However, both groups increased their knowledge significantly. Users described the visualizations as engaging and mentioned that it was more motivating to learn. The animations and visualizations helped them to understand the conceptual operations in the experiment. These results are consistent with previous studies in this area that have shown the potential of animated and interactive visualizations \cite{wolff2000satsim}. It has also been shown that prosocial behavior and sympathy between group members increase in collaborative learning environments \cite{aronson2002building, van2020mediators}. This may be one reason why users did not feel uneasy about exposing their knowledge gaps or were reluctant to speak openly with their team partners. Communication and social interaction in the virtual world offer exciting opportunities for different educational settings \cite{Guetl2011}. Knowing that there is a challenge at the end motivated the users to be more active. We observed that many participants went beyond the tasks and investigated algorithms that were not asked. During the experiment, multi-users felt slightly happier and spent more time on the learning activity. Several multi-user pairs remained in the experiment after all tasks to compete more algorithms against each other in the battle-view. They rated the battle-view as an outstanding positive feature. The overall positive responses regarding the sorting challenges indicate a high acceptance of competitive elements in a collaborative environment. Even the single-users suggested a battle mode where players can compete against each other. This reflects also a high level of acceptance of such competitive elements in the single-user group. Although competitions can increase engagement and have the potential to improve learning outcomes \cite{cerny2011competitive}, there is the risk of losing motivation if one loses the competition. Nevertheless, users felt satisfied, excited, and curious during the experiment. Multi-users felt also less insecure as they were able to support each other. However, losing a competition can lead to negative emotions and reduce enjoyment in the task \cite{vansteenkiste2003competitively}. This can also affect learning success and user acceptance and should be considered in the learning activities. A high user acceptance depends on the system's effectiveness, efficiency, and user satisfaction \cite{brooke2013sus}. While usability was rated equally well, single users were more satisfied with the user interface. This might be due to the added effort required for multi-users to join a server and manage control settings. Users seek a user-friendly system for enjoyable online discussions, connecting with others, and assisting fellow students.

\subsection{Limitations}

The main limitation of our study is the relatively small sample size of 35 participants. A larger and more diverse sample would lead to stronger and more generalized conclusions. The learning outcome was determined by theoretical questions before and after the experiment and did not indicate long-term effects. While all participants had a computer science background, they varied in education level (7 bachelor students, 14 master students, 13 graduated students). Although the single-user group performed better than the multiple-user participants on both tests, this may be because it is more difficult for students to improve when they start at a higher achievement level. Furthermore, the combination of the lab environment with Discord may have influenced the learning experience. One participant from the multi-user group dropped out and did not answer all questions in the post-questionnaire session.
\section{Conclusion}

In conclusion, the findings highlight the transformative potential of collaborative learning complemented by competitive elements to improve student engagement and learning outcomes. The results indicate that students learn more effectively when they work together. The competitive elements increased the students' engagement through a higher level of involvement. It led to a significant improvement in their learning outcomes when they were more involved in the learning process with other students. Users found that they could rely on their partners and had no problems exposing their knowledge gaps. They had also a higher motivation to pass the quiz at the end. A collaborative environment including competitive challenges has shown to be a valuable tool to support students' conceptual understanding. However, overcoming usability challenges is essential for creating an environment that is accepted by the users. Even if the users rated the usability as good, there is still potential for improvement, especially in learning and collaboration. Both user groups requested a more prominent challenge presentation and multi-users asked for an easier way to join the lab environment. Users also criticized that many parts of the experiment were only accessible to the user in control. Therefore, it is preferable to design the learning activities more involving for all users. Nevertheless, the current solution allows educators to demonstrate the concept and then hand over the control to the students. Future studies could focus on the implementation and evaluation in school and classroom settings to find a pedagogical model that is usable for learners and educators. Exploring how the division of responsibility impacts the decisiveness of participants relative to their expertise could offer valuable insights into group dynamics and decision-making across different contexts. 
%
%
%
\bibliographystyle{splncs04}
\bibliography{references}
\end{document}